\pdfoutput=1
\documentclass[11pt, a4paper]{article}

\usepackage{amsthm}
\usepackage{longtable}
\usepackage{amsmath}
\usepackage{amsfonts}
\usepackage{booktabs}
\usepackage{fullpage}
\usepackage{graphicx}
\usepackage{natbib}
\usepackage{mathtools}
\usepackage{color}
\usepackage[left = 2.5cm, right = 2.5cm, bottom = 3cm, top = 2.5cm]{geometry}
\usepackage{enumitem}
\usepackage{multirow}
\usepackage{rotating}
\usepackage{calc}
\usepackage{bm}
\usepackage{authblk}
\usepackage{tikz}
\usepackage{dcolumn}

\usepackage{setspace}
\usepackage{pdfpages}
\usepackage{xr-hyper}
\usepackage{hyperref}
\usepackage{algorithm}
\usepackage{algpseudocode}

\newtheoremstyle{example}
{3pt} 
{3pt} 
{} 
{0\parindent} 
{\bf}
{:} 
{.5em} 
{} 
\newtheoremstyle{theorem}
{3pt} 
{3pt} 
{\em} 
{0\parindent} 
{\bf}
{:} 
{.5em} 
{} 
\theoremstyle{example} 
\theoremstyle{theorem} 

\def\expect{{\mathop{\rm E}}}
\def\var{{\mathop{\rm var}}}

\title{Bounded-memory adjusted scores estimation in generalized linear models with large data sets}

\author[1]{Patrick Zietkiewicz\thanks{patrick.zietkiewicz@warwick.ac.uk}}
\author[1]{Ioannis Kosmidis\thanks{ioannis.kosmidis@warwick.ac.uk}}

\affil[1]{Department of Statistics, University of Warwick \authorcr Gibbet Hill Road, Coventry, CV4 7AL, UK}

\begin{document}

\maketitle

\begin{abstract}
  The widespread use of maximum Jeffreys'-prior penalized likelihood
  in binomial-response generalized linear models, and in logistic
  regression, in particular, are supported by the results of Kosmidis
  and Firth (2021, Biometrika), who show that the resulting estimates
  are always finite-valued, even in cases where the maximum likelihood
  estimates are not, which is a practical issue regardless of the size
  of the data set. In logistic regression, the implied adjusted score
  equations are formally bias-reducing in asymptotic frameworks with a
  fixed number of parameters and appear to deliver a substantial
  reduction in the persistent bias of the maximum likelihood estimator
  in high-dimensional settings where the number of parameters grows
  asymptotically as a proportion of the number of observations. In
  this work, we develop and present two new variants of iteratively
  reweighted least squares for estimating generalized linear models
  with adjusted score equations for mean bias reduction and
  maximization of the likelihood penalized by a positive power of the
  Jeffreys-prior penalty, which eliminate the requirement of storing
  $O(n)$ quantities in memory, and can operate with data sets that
  exceed computer memory or even hard drive capacity. We achieve that
  through incremental QR decompositions, which enable IWLS iterations
  to have access only to data chunks of predetermined size. Both
  procedures can also be readily adapted to fit generalized linear
  models when distinct parts of the data is stored across different
  sites and, due to privacy concerns, cannot be fully transferred
  across sites.  We assess the procedures through a real-data
  application with millions of observations.
\end{abstract}

\section{Introduction}

\subsection{Boundary maximum likelihood estimates}

\citet{albert+anderson:1984} show that, in logistic regression, data
separation is a necessary and sufficient condition for the maximum
likelihood (ML) estimate to have at least one infinite component. Data
separation occurs when there is a linear combination of covariates
that perfectly predicts the response values, and results in the
log-likelihood having an asymptote in some direction of the parameter
space. When separation occurs, ML fitting procedures typically result
in apparently finite estimates and estimated standard errors, which
are mere artefacts of the numerical optimization procedure stopping
prematurely by meeting the optimizer's convergence criteria, or
reaching the maximum allowed number of iterations. If infinite
estimates go undetected, then inference about the model parameters can
lead to misleading conclusions. \citet{mansournia+etal:2018} is a
recent overview of the practical issues associated with infinite
estimates in logistic regression. Infinite ML estimates can also occur
for other binomial-response regression models. 

The detection of separation and, more generally, identification of
which components of the maximum likelihood estimate are infinite can
be done prior to fitting the model by solving appropriate linear
programs. For example, the \texttt{detectseparation} R package
\citep{detectseparation} provides methods that implement the linear
programs in \citet{konis:2007} for many binomial-response generalized
linear models (GLMs), and the linear program in
\citet{schwendinger+etal:2021} for log-binomial models for relative
risk regression. Such procedures, however, while helpful to identify
issues with ML, they provide no constructive information about what to
do when infinite ML estimates are encountered.

It is important to note that infinite estimates occur for both small
and large data sets. For an illustration, see
Section~\ref{sec:air2000}, where infinite estimates are observed for a
probit regression model fit on a data set with about 5.5 million
observations and 37 covariates. Furthermore, for logistic regression
models with $n$ observations and $p$ covariates, where
$p/n \to \kappa \in (0, 1)$, \citet{candes+sur:2020} prove that for
model matrices generated from particular distributions, there are
combinations of $\kappa$ and the variance of the linear predictor for
which the ML estimate has infinite components with probability one as
$p/n \to \kappa$, regardless of the specific values of $p$ or $n$.

The linear programs for detecting infinite estimates can be slow for
large model matrices, and infeasible in their vanilla implementations
if the model matrix and response vector do not fit in computer
memory. For example, solving the \citet{konis:2007} linear program for
the model of Section~\ref{sec:air2000} with the default settings of
\texttt{detect\_infinite\_estimates()} method of the {\tt
  detectseparation} R package did not return a result even after 12
hours of computation on a 2021 MacBook Pro with an Apple M1 Max chip
and 64 GB RAM.

\subsection{Alternative estimators}

To summarize, on one hand, ML estimation for binomial-response GLMs
may lead to arbitrarily large estimates and estimated standard errors,
which are in reality infinite, and if they go undetected can cause
havoc to standard inferential procedures. On the other hand, linear
programs for detecting the occurrence of infinite ML estimates prior
to fitting are not constructive about the consequences of having
infinite estimates in the model fit or the impact this may have on
other parameters in the model, and have either long run times for
large data sets or their standard implementations are infeasible for
data sets that do not fit in computer memory.

For those reasons, researchers typically resort to alternative
estimators, which in many settings have the same optimal properties as
the ML estimator has, and are guaranteed to be finite even in cases
where the ML estimator is not, regardless of the data set or its
dimension. For example, \citet{kosmidis+firth:2021} show that
penalization of the likelihood by Jeffreys' invariant prior, or by any
positive power thereof, always produces finite-valued maximum
penalized likelihood estimates in a broad class of binomial-response
GLMs, only under the assumption that the model matrix is of full
rank. Another approach that has been found to deliver finite estimates
for many binomial-response GLMs are the adjusted score equations for
mean bias reduction (mBR) in \citet{firth:1993}, which, for fixed $p$,
result in estimators with mean bias of smaller asymptotic order than
what the ML estimator has in general. To date, there is only empirical
evidence that the reduced-bias estimator of \citet{firth:1993} has
always finite components for binomial-response GLMs with arbitrary
link functions. For fixed-$p$ logistic regression, the bias-reducing
adjusted score functions end up being the gradient of the logarithm of
the likelihood penalized by Jeffreys' invariant prior. Hence, in that
case, the results in \citet{kosmidis+firth:2021} show that the
reduced-bias estimators have also always finite components.

Both \citet{sur+candes:2019} and \citet{kosmidis+firth:2021} also
illustrate that maximum Jeffreys'-prior penalized likelihood (mJPL)
results in a substantial reduction in the persistent bias of the ML
estimator in high-dimensional logistic regression problems with
$p/n \rightarrow \kappa \in (0, 1)$, when the ML estimator has finite
components. Those results underpin the increasingly widespread use of
mBR and similarly penalized likelihood estimation in binomial
regression models in many applied fields.

\subsection{Our contribution}

The ML estimates for GLMs can be computed through iterative reweighted
least squares (IWLS; \citealt{green:1984}). Because each element of
the working variates vector for the IWLS iteration depends on the
current value of the parameters and the corresponding observation, ML
estimation can be performed even with larger-than-memory data sets
using incremental QR decompositions as in \citet{miller:1992}. The
incremental IWLS for ML is implemented in the \texttt{biglm} R package
\citep{lumley:2020} for all GLMs.

In this work, we begin by presenting a unifying IWLS procedure for mBR
and mJPL in GLMs through a modification of the ML working
variates. Unlike the working variates for ML, the modified working
variates involve quantities, like the leverage, that depend on the
whole data set
\citep[Section~2.3]{kosmidis+kennepagui+sartori:2020}. As a result,
IWLS through incremental QR decompositions is not directly
possible. To overcome this difficulty, we present and analyze the
properties of two variants of IWLS for solving the mBR and mJPL
adjusted score equations, which only require access to data blocks of
fixed size the user can specify in light of any local memory
constraints. We should emphasize that the block size can be arbitrary
and the resulting estimates are invariant to its choice. As a result,
the two IWLS variants eliminate the requirement of keeping $O(n)$
quantities in memory, opening the door for fitting GLMs with adjusted
score equations and penalized likelihood methods on data sets that are
larger than local memory or hard drive capacity, and are stored in a
remote database. The procedures operate with either one- or two-passes
through the data set per iteration, and return the mBR and mJPL
estimates.

Importantly, both procedures can be readily adapted to fit GLMs when
distinct parts of the data is stored across different sites and, due
to privacy concerns, cannot be fully transferred across sites. In
light of the results in \citet{kosmidis+firth:2021}, such adaptations
provide guarantees of stability of all numerical and inferential
procedures even when the data set is separated and infinite ML
estimates occur, in settings where checking for infinite estimates is
not feasible with existing algorithms.

\subsection{Organization}

Section~\ref{sec:glms} sets notation by introducing GLMs and their ML
estimation using IWLS, and details how that estimation can be done
using only data chunks of fixed size through incremental QR
decompositions. Section~\ref{sec:br_mjpl} presents the adjusted score
equations for mBR and mJPL in GLMs, and introduces and analyzes the
memory and computation complexity of two variants of IWLS that operate
in a chunk-wise manner, avoiding the memory issues that vanilla IWLS
implementations can encounter. Section~\ref{sec:air2000} applies the
algorithms for the modelling of the probability of a diverted flight
using probit regression, based on data from all $5\,683\,047$ commercial
flights within the USA in 2000. Section~\ref{sec:privacy} presents
that adaptations of the two variants in settings where distinct parts
of the data is stored across different sites and, due to privacy
concerns, cannot be fully transferred across sites. Finally,
Section~\ref{sec:concluding} provides discussion and concluding
remarks.

\section{Generalized linear models}
\label{sec:glms}
\subsection{Model}

Suppose that $y_1, \ldots, y_n$ are observations on random variables
$Y_1, \ldots, Y_n$ that are conditionally independent given
$x_1, \ldots, x_n$, where $x_{i}$ is a $p$-vector of covariates. A GLM
\citep{mccullagh+nelder:1989} assumes that, conditionally on $x_i$,
$Y_i$ has an exponential family distribution with probability density
or mass function of the form
\[
L(\mu_i, \phi; y) = \exp\left\{\frac{y \theta_i - b(\theta_i) - c_1(y)}{\phi/m_i} - \frac{1}{2}a\left(-\frac{m_i}{\phi}\right) + c_2(y) \right\}
\]
for some sufficiently smooth functions $b(.)$, $c_1(.)$, $a(.)$ and
$c_2(.)$, and fixed observation weights $m_1, \ldots, m_n$, where $\theta_i$
is the natural parameter. The expected value and the variance of $Y_i$
are then
\begin{align*}
      \expect(Y_i) & = \mu_i =  b'(\theta_i) \\
      \var(Y_i) & = \frac{\phi}{m_i}b''(\theta_i) = \frac{\phi}{m_i}V(\mu_i) \, .
\end{align*}
Hence, the parameter $\phi$ is a dispersion parameter. The mean
$\mu_i$ is linked to a linear predictor $\eta_i$ through a a monotone,
sufficiently smooth link function $g(\mu_i) = \eta_i$ with
\begin{equation}
  \label{linear_predictor}
\eta_i = \sum_{t=1}^p \beta_t x_{it}
\end{equation}
where $x_{it}$ can be thought of as the $(i,t)$th component of a model
matrix $X$, assumed to be of full rank, and
$\beta = (\beta_1, \ldots, \beta_p)^\top$. An intercept parameter is
customarily included in the linear predictor, in which case
$x_{i1} = 1$ for all $i \in \{1, \ldots, n\}$.

\subsection{Likelihood, score functions and information}
The log-likelihood function for a GLM is
$\ell(\beta) = \sum_{i = 1}^n \log L(F(\eta_i), \phi; y)$, where
$F(\cdot) = g^{-1}(.)$ is the inverse of the link function, and
$\eta_i$ is as in~(\ref{linear_predictor}). Temporarily suppressing
the dependence of the various quantities on the model parameters and
the data, the derivatives of the log-likelihood function with respect
to the components of $\beta$ and $\phi$ are
\begin{equation}
  \label{eq:scores}
  s_\beta = \frac{1}{\phi}X^TW (z - \eta) \quad \text{and} \quad
  s_\phi = \frac{1}{2\phi^2}\sum_{i = 1}^n (q_i - \rho_i) \, ,
\end{equation}
respectively, where $z = \eta + D^{-1}(y - \mu)$ is the vector of
working variates for ML, $\eta = X \beta$,
$y = (y_1, \ldots, y_n)^\top$, $\mu = (\mu_1, \ldots, \mu_n)^\top$,
$W = {\rm diag}\left\{w_1, \ldots, w_n\right\}$ and
$D = {\rm diag}\left\{d_1, \ldots, d_n\right\}$, with
$w_i = m_i d_i^2/v_i$ being the $i$th working weight, and
$d_i = d\mu_i/d\eta_i = f(\eta_i)$, $v_i = V(\mu_i)$. Furthermore,
$q_i = -2 m_i \{y_i\theta_i - b(\theta_i) - c_1(y_i)\}$ and
$\rho_i = m_i a'_i$ are the $i$th deviance residual and its
expectation, respectively, with $a'_i = a'(-m_i/\phi)$, where
$a'(u) = d a(u)/d u$.

The ML estimators $\hat\beta$ and $\hat\phi$ can be found by solution
of the score equations $s_{\beta} = 0_p$ and $s_\phi = 0$, where $0_p$
is a $p$-vector of zeros. \citet{wedderburn:1976} derives necessary
and sufficient conditions for the existence and uniqueness of the ML
estimator. Given that the dispersion parameter $\phi$ appears in the
expression for $s_{\beta}$ in~(\ref{eq:scores}) only multiplicatively,
the ML estimate of $\beta$ can be computed without knowledge of the
value of $\phi$. This fact is exploited in popular software like the
\texttt{glm.fit()} function in R \citep{rproject}. The $j$th iteration
of IWLS updates the current iterate $\beta^{(j)}$ by solving the
weighted least squares problem
\begin{equation}
  \label{eq:iwls}
  \beta^{(j+1)} := \left(X^\top W^{(j)} X\right)^{-1} X^\top W^{(j)}z^{(j)}\,,
\end{equation}
where the superscript $(j)$ indicates evaluation at $\beta^{(j)}$
\citep{green:1984}. The updated $\beta$ from~(\ref{eq:iwls}) is equal
to that from the Fisher scoring step
$\beta^{(j)} + \{i_{\beta\beta}^{(j)}\}^{-1} s_{\beta}^{(j)}$ where
$i_{\beta\beta}$ is the $(\beta,\beta)$ block of the expected
information matrix about $\beta$ and $\phi$
\begin{equation}
  \label{eq:information}
\left[
\begin{array}{cc}
i_{\beta\beta} & 0_p \\
0_p^\top & i_{\phi\phi}
\end{array}
\right]
=
\left[
\begin{array}{cc}
\frac{1}{\phi} X^\top W X & 0_p \\
0_p^\top & \frac{1}{2\phi^4}\sum_{i = 1}^n m_i^2 a''_i
\end{array}
\right]\,,
\end{equation}
with $a''_i = a''(-m_i/\phi)$, where $a''(u) = d^2 a(u)/d u^2$.

The weighted least squares problem in~(\ref{eq:iwls}) is typically
solved either through the method of normal equations, which involves a
Cholesky decomposition of $X^\top W^{(j)} X$, or through the QR
decomposition of $(W^{(j)})^{1/2} X$ \citep[see,
e.g.][Section~5.3]{golub+vanloan:2013}. Despite that the QR approach
requires more computations than the method of normal equations, the
former is more appealing in applications and the default choice in
popular least squares software because it can solve, in a numerically
stable manner, a wider class of least squares problems. In particular,
the method of normal equations can numerically break down even when
$(W^{(j)})^{1/2} X$ is not particularly close to being numerically
rank deficient, while the QR approach solves a ``nearby'' least
squares problem \citep[see,][Section~5.3 for an extensive
analysis]{golub+vanloan:2013}.

ML estimation of $\phi$ can then take place by solving $s_\phi = 0$
after evaluating $q_i$ at the ML estimates $\hat\beta$. This can be
done through the Fisher scoring iteration
\begin{equation}
  \label{eq:fs_phi}
  \phi^{(j+1)} := \phi^{(j)} \left\{ 1 + \phi^{(j)} \frac{\sum_{i = 1}^n (\hat{q}_i - \rho_i^{(j)})}{\sum_{i = 1}^n m_i^2 a_i''^{(j)}} \right\} \,,
\end{equation}
where $\hat{q}_i$ is $q_i$ evaluated at
$\hat\beta$. \citet{mccullagh+nelder:1989} recommend against
estimating $\phi$ using ML, and instead propose the moment estimator
$(n - p)^{-1} \sum_{i = 1}^n \hat{w}_i (\hat{z}_i - \hat\eta_i)^2$
where $\hat{w}_i$, $\hat{z}_i$, and $\hat{\eta_i}$ are $w_i$, $z_i$,
and $\eta_i$, respectively, evaluated at $\hat\beta$. The moment
estimator of $\phi$ is considered to have less bias that the ML
estimator.

\subsection{Bounded-memory procedures for least squares}
\label{sec:bmem_al}

\citet{miller:1992} proposes a procedure to solve least squares
problems using the QR approach, which does not require keeping the
full model matrix $X$ and response vector $y$ in memory, and operates
by sequentially bringing only a fixed-size chunk of data in
memory. This is particularly useful in getting a numerically stable
least squares solution in cases where the model matrix has too many
rows to fit in memory. We briefly review the method in
\citet{miller:1992}.

Consider the least squares problem $A \psi = b$, where $A$ is an
$n \times p$ matrix, with $n > p$, and $b$ is a $n$-dimensional
vector. The least squares solution $\hat\psi$ for $\psi$ can be found
by first computing the QR decomposition
\[
  A = QR = \left[
    \begin{array}{cc}
      \bar{Q} & Q^*
    \end{array}
  \right]
  \left[
    \begin{array}{c}
      \bar{R} \\
      0_{(n - p) \times p}
    \end{array}
  \right]\,,
\]
where $\bar{Q}$ and $Q^*$ are $n \times p$ and $n \times (n - p)$
matrices, respectively, $Q$ is an orthogonal matrix
(i.e.~$Q^\top = Q^{-1}$), $\bar{R}$ is a $p \times p$ upper triangular
matrix, and $0_{u \times v}$ denotes a $u \times v$ matrix of
zeros. Then, $\hat\psi$ is found by using back-substitution to solve
$\bar{R} \hat\psi = \bar{b}$, where
$\bar{b} = \bar{Q}^\top b = \{\bar{R}^{-1}\}^\top A^\top b$. To
describe the proposal in \citet{miller:1992}, we partition the rows of
$A$ into $K$ chunks, with the $k$th chunk having $c_k \le n$ rows
$(k = 1, \ldots K)$. Let $d_k = \sum_{j = 1}^k c_k$. So $n = d_K$. We
use the same partitioning for the elements of $b$.  Denote by $A_{:k}$
and $b_{:k}$ the $k$th chunks of $A$ and $b$, respectively, and by
$A_{1:k}$ the first $k$ chunks of $A$. Suppose that the QR
decomposition of $A_{1:k}$ is $Q_{:k} R_{:k}$. Then, in light of a new
chunk $A_{:k + 1}$, the QR decomposition can be updated as
\begin{equation}
  \label{eq:qr-update}
  \left[
    \begin{array}{c}
      A_{1:k} \\
      A_{:k + 1}
    \end{array}
  \right]   =
  \underbrace{\left[
    \begin{array}{ccc}
      \bar{Q}_{:k} & 0 & Q_{:k}^* \\
      0 & I & 0
    \end{array}
  \right]
  G_1 \cdots G_{cp}}_{Q_{:k + 1}} \overbrace{G_{cp}^\top \cdots G_{1}^\top
  \left[
    \begin{array}{c}
      \bar{R}_{:k} \\
      A_{:k + 1} \\
      0
    \end{array}
  \right]}^{R_{:k+1}}\,,
\end{equation}
where $G_1, \ldots, G_{cp}$ is the set of the Givens rotation matrices
required to eliminate (i.e.~set to zero) the elements of $A_{:k+1}$ in
the right hand side (see \citealt[Section 5.1.8]{golub+vanloan:2013},
or Section~\ref{sec:supp_givens} in the Supplementary Material
document, for the definition and properties of Givens rotation
matrices). For keeping the notation simple, zero and identity matrices have
temporarily been denoted as $0$ and $I$, with their dimension being
implicitly understood, so that the product of blocked matrices is
well-defined.

By the orthogonality of Givens rotation matrices, $G_1, \ldots, G_{cp} G_{cp}^\top, \ldots, G_{1}^\top = I_{d_{k + 1}}$, so, taking products of blocked matrices, equation~(\ref{eq:qr-update}) simply states that $A_{1:k} = \bar{Q}_{:k} \bar{R}_{:k}$ and that $A_{:k+1} = A_{:k+1}$. However, note that 
$$
Q_{:k + 1} = \left[
    \begin{array}{ccc}
      \bar{Q}_{:k} & 0 & Q_{:k}^* \\
      0 & I & 0
    \end{array}
  \right]
  G_1 \cdots G_{cp}
$$
is orthogonal, as the product of orthogonal matrices, and that
$$
R_{:k+1} = G_{cp}^\top \cdots G_{1}^\top
  \left[
    \begin{array}{c}
      \bar{R}_{:k} \\
      A_{:k + 1} \\
      0
    \end{array}
  \right] \,,
$$
has its first $p$ rows, $\bar{R}_{:k+1}$, forming an upper triangular
matrix, and all of its other elements are zero, due to the Givens
rotations. So, $Q_{:k + 1} R_{:k + 1}$ is a valid QR decomposition.

The only values that are needed for computing the least squares
estimates $\hat\psi$ are $\bar{R}$ and $\bar{b}$. Hence, their current
values are the only quantities that need to be kept in
memory. Furthermore, since storage of the current value of $Q$ is not
necessary, and since a Givens rotation acts only on two rows of the
matrix it pre-multiplies (see Section~\ref{sec:supp_givens} in the
Supplementary Material document), there is no need to form any $O(n)$
objects during the update. This is useful for large $n$, as the need
of having an $n \times p$ matrix in memory ($A$) is replaced by
keeping $p(p + 3)/2$ real numbers in memory at any given
time. Without loss of generality, in what follows we assume that
$c_1 = c_2 = \ldots = c_{K-1} = c \le n$, and $c_K = n - (K - 1) c$.

Procedure {\tt updateQR} in Algorithm~\ref{algo:supp_updateQR} of the
Supplementary Material document updates the current value of $\bar{R}$
and $\bar{b}$, given a chunk of $A$ and the corresponding chunk of
$b$, using \citet[Algorithm~AS274.1]{miller:1992}, which does the
update for a single observation. Procedure {\tt incrLS} in
Algorithm~\ref{algo:supp_incrLS} uses {\tt updateQR} to compute the
least squares estimates of $\psi$ by incrementally updating the value
of $\bar{R}$ and $\bar{b}$, using one chunk of $c$ observations at a
time (with the last chunk containing $c$ or fewer observations).

\subsection{Iteratively re-weighted least squares in chunks}
\label{sec:chunkwiseIWLS}
The approach of \citet{miller:1992} can be used for~(\ref{eq:iwls}) by
replacing $A$ by $(W^{(j)})^{1/2} X$ and $b$ by
$(W^{(j)})^{1/2}z^{(j)}$. This is possible because the diagonal
entries of $W$ and the components of $z$ only depend on the
corresponding components of $\eta = X \beta$, and hence, can be
computed in a chunk-wise manner. Specifically, the $k$th chunk of the
diagonal of $W^{(j)}$, and the $k$th chunk of $z^{(j)}$ can be
computed by just bringing in memory $X_{:k}$, $Y_{:k}$, and $m_{:k}$,
computing the current value of the $k$th chunk of the linear predictor
as $X_{:k} \beta^{(j)}$, and using that to compute the $k$th chunks of
$\mu^{(j)}$, $d^{(j)}$ and
$v^{(j)}$. Algorithm~\ref{algo:supp_incrIWLS} in the Supplementary
Material document provides pseudo-code for the above process, which is
also implemented in the \texttt{biglm} R package \citep{lumley:2020}.

\section{Bias reduction and maximum penalized likelihood}
\label{sec:br_mjpl}

\subsection{Adjusted score equations}

Consider the adjusted score equations
\begin{align}
  \label{eq:adj_eq_beta}
  0_p & = \frac{1}{\phi} X^\top W \left\{z - \eta + \phi H (b_1 \xi + b_2 \lambda) \right\}\,, \\
  \label{eq:adj_eq_phi}
  0 & = \frac{1}{2\phi^2} \sum_{i = 1}^n (q_i - \rho_i) + c_1 \left( \frac{p - 2}{2 \phi} + \frac{\sum_{i = 1}^n m_i^3 a_i'''}{2\phi^2\sum_{i = 1}^n m_i^2 a_i''} \right) \, ,
\end{align}
where $0_p$ is a $p$-vector of zeros,
$\xi = (\xi_1, \ldots, \xi_n)^\top$, and
$\lambda = (\lambda_1, \ldots, \lambda_n)^\top$ with
\begin{align*}
  \xi_i = \frac{d_i'}{2 d_i w_i}\,, \quad \text{and} \quad
  \lambda_i = \frac{1}{2} \left\{\frac{d_i'}{d_i w_i} - \frac{v_i'}{m_i d_i}\right\} \,, \quad  
\end{align*}
and $a'''_i = a'''(-m_i/\phi)$, where $a'''(u) = d^3 a(u)/d u^3$,
$d_i' = d^2\mu_i/d\eta_i^2 = f'(\eta_i)$, and
$v'_i = d V(\mu_i) / d\mu_i$. In the above expressions, $H$ is a
diagonal matrix, whose diagonal is the diagonal of the ``hat'' matrix
$X (X^\top W X)^{-1} X^\top W$. The ML estimators of $\beta$ and
$\phi$ are obtained by solving~\eqref{eq:adj_eq_beta}
and~\eqref{eq:adj_eq_phi} with $b_1 = b_2 = c_1 =
0$. \citet{kosmidis+kennepagui+sartori:2020} show that mBR estimators
of $\beta$ and $\phi$ can be obtained by
solving~\eqref{eq:adj_eq_beta} and~\eqref{eq:adj_eq_phi} with
$b_1 = 1$, $b_2 = 0$, and $c_1 = 1$. On the other hand, direct
differentiation of the Jeffreys'-prior penalized likelihood shows that
the mJPL estimators for
$\beta = \arg\max \{ \ell(\beta) + \log|X^\top W X| / 2 \}$ in
\citet{kosmidis+firth:2021} can be obtained by
solving~\eqref{eq:adj_eq_beta} for $b_1 = b_2 = 1$.

For binomial-response GLMs, \citet{kosmidis+firth:2021} show that the
mJPL estimate of $\beta$ has always finite components for a wide range
of well-used link functions, including logit, probit and complementary
log-log, even in cases where the ML estimate has infinite
components. The finiteness property of the mJPL estimator is
attractive for applied work, because it ensures the stability of all
numerical and inferential procedures, even when the data set is
separated and infinite ML estimates occur \citep[see,][for a recent
review of the practical issues associated with infinite estimates in
logistic regression]{mansournia+etal:2018}. The same holds for any
power $t > 0$ of the Jeffreys' prior, in which case
$\lambda_i = t \{d_i'/ (d_i w_i) - v_i'/(m_i d_i)\} / 2$
in~\eqref{eq:adj_eq_beta}.

The mJPL estimators, though, do not necessarily have better asymptotic
bias properties than the ML estimator for all GLMs . For GLMs that are
full exponential families, such as logistic regression for binomial
data and log linear models for Poisson counts, the mJPL estimators
with $t = 1$ and the mBR estimators coincide. This has been shown
in~\citet{firth:1993}, and is also immediately apparent
by~\eqref{eq:adj_eq_beta}. For canonical links $d_i = v_i$,
$w_i = m_i d_i$, and, by the chain rule $v_i' = d_i' / d_i$, hence
$\lambda_i = 0$. Thus, the mBR estimator for logistic regression not
only has bias of smaller asymptotic order than what the ML estimator
generally has, but their components also always take finite values.

We should note that the first-order asymptotic properties expected by
the ML, mJPL and mBR estimators of $\beta$ and $\phi$ are preserved
for any combination of $b_1$, $b_2$ and $c_1$
in~\eqref{eq:adj_eq_beta} and~\eqref{eq:adj_eq_phi}. For example, mBR
estimators of $\beta$ can be obtained for $b_1 = 1$, $b_2 = 0$ and
$c_1 = 0$, effectively mixing mBR for $\beta$ with ML for $\phi$. This
is due to the orthogonality of $\beta$ and $\phi$
\citep{cox+reid:1987}; the detailed argument is a direct extension of
the argument in \citet[Section~4]{kosmidis+kennepagui+sartori:2020}
for mixing adjusted score equations to get mBR for $\beta$ and
estimators of $\phi$ that have smaller median bias. Another option is
to mix the adjusted score equations~\eqref{eq:adj_eq_beta} for $\beta$
with the estimating function
\begin{equation}
  \label{eq:phi_mom}
  \phi = \frac{1}{n - p} \sum_{i = 1}^n w_i (z_i - \eta_i)^2 \, ,
\end{equation}
which gives rise to the moment estimator of $\phi$, once $\mu_i$ is
evaluated at an estimator for $\beta$.

\subsection{Iteratively re-weighted least squares for solving adjusted score equations}
\label{sec:iwls_adj}
Using similar arguments as for ML estimation, we can define the
following IWLS update to compute the ML, the mBR, and the mJPL
estimators of $\beta$ depending on the value of $b_1$, $b_2$
in~(\ref{eq:adj_eq_beta}). That update has the form
\begin{equation}
  \label{eq:iwlsadj}
  \beta^{(j+1)} := \left(X^\top W^{(j)} X\right)^{-1} X^\top W^{(j)}\left(z^{(j)} + \phi^{(j)} H^{(j)} \kappa^{(j)}\right)\,,
\end{equation}
where $\kappa = b_1 \xi + b_2 \lambda$. The value of $\phi$, which is required for implementing mBR and mJPL,
can be found via the quasi-Fisher scoring step for
solving~(\ref{eq:adj_eq_phi}), which using~(\ref{eq:information}),
takes the form
\begin{equation}
  \label{eq:fs_phi_adj}
  \phi^{(j+1)} := \phi^{(j)} \left[ 1 + \phi^{(j)} \frac{\sum_{i = 1}^n (q_i^{(j)} - \rho_i^{(j)})}{\sum_{i = 1}^n m_i^2 a_i''^{(j)}} +
    c_1\phi^{(j)} \left\{\frac{\sum_{i = 1}^n m_i^3 a_i'''^{(j)}}{(\sum_{i = 1}^n m_i^2 a_i''^{(j)})^2} +
    \phi^{(j)} \frac{p-2}{\sum_{i = 1}^n m_i^2 a_i''^{(j)}} \right\} \right]
\end{equation}
The candidate value for the dispersion parameter when one of $b_1$ or
$b_2$ is non-zero, can be computed using~(\ref{eq:phi_mom}) at
$\beta^{(j)}$ in every
iteration as
\begin{equation}
  \label{eq:phi_mom_update}
  \phi^{(j + 1)} = \frac{1}{n - p} \sum_{i = 1}^n w_i^{(j)} (z_i^{(j)} - \eta_i^{(j)})^2 \, .
\end{equation}
\citet{kosmidis+kennepagui+sartori:2020} have already
derived the special case of~(\ref{eq:iwlsadj})
and~(\ref{eq:fs_phi_adj}) for mBR estimation, that is for $b_1 = 1$,
$b_2 = 0$ and $c_1 = 1$. Convergence can be declared if
$\lVert \beta^{(j + 1)} - \beta^{(j)} \rVert_\infty < \epsilon$ and
$\lVert \phi^{(j' + 1)} - \phi^{(j)} \rVert_\infty < \epsilon$ for
some small $\epsilon > 0$.

\subsection{Adjusted score estimation in chunks}

When either~(\ref{eq:fs_phi}) or~(\ref{eq:phi_mom_update}) is used,
the updates for $\phi$ can be performed in a chunkwise manner once
$\beta^{(j)}$ has been computed. In fact, use
of~(\ref{eq:phi_mom_update}) has the advantage of requiring only
$w^{(j)}$ and $z^{(j)}$, which have already been computed after
performing the IWLS update~(\ref{eq:adj_eq_beta}).

Nevertheless, the IWLS update~(\ref{eq:iwlsadj}) for $\beta$ cannot be
readily performed in a chunkwise manner because the diagonal of $H$
cannot be computed at $\beta^{(j)}$ by just bringing in memory chunks
of $X$, $Y$ and $m$, and the current estimates. In particular, the
$i$th diagonal element of $H$ is
$h_i = x_i^\top (X^\top W X)^{-1} x_i w_i$, and, hence, its
computation requires the inverse of the expected information matrix
$X^\top W X$. In what follows, we present two alternatives for
computing adjusted score estimates for $\beta$ in a chunkwise manner.

\subsection{Two-pass implementation}
\label{sec:two-pass}

\begin{figure}[t]
  \caption{Demonstration of the IWLS update~(\ref{eq:iwlsadj}). All
    quantities in the figure should be understood as being
    pre-multiplied by $W^{1/2}$ and evaluated at $\beta^{(j)}$ and
    $\phi^{(j)}$. The left figure shows the addition of $\phi H\kappa$
    to the ML working variates vector $z$, and the
    subsequent projection onto $\mathcal{C}$ (the column space of
    $W^{1/2} X$) that gives the updated value for the mBR or mJPL
    estimates $\tilde{\beta}$. The right figure shows the projection
    of $z$, and the subsequent projection of $\phi H \kappa$ onto
    $\mathcal{C}$, and the addition of the projected vectors that
    gives the updated value for the mBR or mJPL estimates
    $\tilde{\beta}$.}
  \begin{center}
    \includegraphics[width = \textwidth]{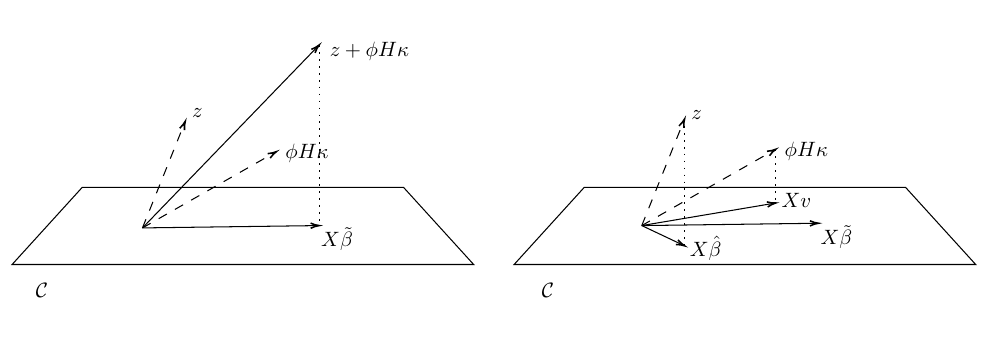}
  \end{center}
  \label{fig:iwls_br_figure}
\end{figure}

The direct way to make the update~(\ref{eq:adj_eq_beta}) for $\beta$
possible in a chunk-wise manner comes from the considering the
projection that is being made. The left plot of
Figure~\ref{fig:iwls_br_figure} shows how
update~(\ref{eq:adj_eq_beta}) projects the current value of the vector
$W^{1/2}(z + \phi H \kappa)$ onto the column space of the current
value of $W^{1/2} X$.

The right of Figure~\ref{fig:iwls_br_figure}, then shows how we can
achieve exactly the same projection in two passes through the data: i)
project the current value of $W^{1/2}z$ onto the column space of the
current value of $W^{1/2} X$ using QR decomposition, and ii) project
the current value of $\phi W^{1/2} H \kappa$ onto the column space of
the current value of $W^{1/2} X$. Adding up the coefficient vectors
from the two projections returns the required updated value of
$\beta$. Section~\ref{sec:chunkwiseIWLS} describes how the projection
in step i) can be done in a chunkwise manner. Then, given that the
current value of $\bar{R}$ is available after the completion of the
incremental QR decomposition from the first pass through the data, the
current value of $h_1, \ldots, h_n$ in step ii) can be computed in a
chunkwise manner through another pass through the data.

Algorithm~\ref{algo:supp_incrIWLS2} in the Supplementary Material
document provides pseudo-code for the two-pass implementation of
adjusted score estimation of $\beta$, which is also implemented in the
port of the \texttt{biglm} R package provided in the Supplementary
Material.

\subsection{One-pass implementation}
\label{sec:one-pass}
An alternative way to make the update~(\ref{eq:adj_eq_beta}) for
$\beta$ possible in a chunk-wise manner is to change the IWLS update to
\begin{equation}
  \label{eq:iwls1pass}
  \beta^{(j+1)} := \left(X^\top W^{(j)} X\right)^{-1} X^\top W^{(j)}(z^{(j)} + \phi^{(j)} H^{(j - 1)} \kappa^{(j)})\, .
\end{equation}
Iteration~(\ref{eq:iwls1pass}) has the correct stationary point, that
is the solution of the adjusted score
equations~(\ref{eq:adj_eq_beta}), and can be performed in a chunkwise
manner following the descriptions in Section~\ref{sec:chunkwiseIWLS}.
This is because the $i$th diagonal element of $H^{(j - 1)}$,
$h_i = x_i^\top (X^\top W^{(j - 1)} X)^{-1} x_i w_i^{(j - 1)}$,
depends on the weight $w_i^{(j - 1)}$ at the previous value of
$\beta$, which can be recomputed using knowledge of $\beta^{(j - 1)}$
and $x_i$ only, and the inverse of the expected information
$(X^\top W^{(j - 1)} X)^{-1}$ from the previous iteration, which can
be computed using $\bar{R}^{(j-1)}$ that is available from the
incremental QR decomposition at the $(j-1)$th iteration.

In addition to the starting value for $\beta^{(0)}$ that
iteration~(\ref{eq:iwls}) requires, iteration~(\ref{eq:iwls1pass})
also requires starting values for $h_i$; a good starting value is
$h_{i}^{(0)} = p / n$, which corresponds to a balanced model matrix
$X$, or the value of $h_i$ by letting the first iteration be an IWLS
step for ML estimation.

Algorithm~\ref{algo:supp_incrIWLS1} in the Supplementary Material
document provides pseudo-code for the one-pass implementation of
adjusted score estimation of $\beta$, which is also implemented in the
port of the \texttt{biglm} R package provided in the Supplementary
Material. 

\subsection{Memory requirements and computational complexity}
\label{sec:mem_cpu}

The direct implementation of IWLS for mBR and mJPL, as done in popular
R packages such as \texttt{brglm2} \citep{brglm2} and \texttt{logistf}
\citep{logistf}, requires $O(n p + p^2)$ memory, as ML does. On the
other hand, the chunkwise implementations require $O(c p + p^2)$
memory, where $c$ is the user-specified chunk size as in
Section~\ref{sec:bmem_al}. The computational cost of all
implementations using the QR decomposition remains at
$O(n p^2 + p^3)$.

The two-pass implementation has almost twice the iteration cost of the
one-pass implementation, because two passes through all observations
is required per IWLS iteration. However, the two-pass implementation
reproduces exactly iteration~(\ref{eq:iwlsadj}) where the adjusted
score, rather that just part of it as in~(\ref{eq:iwls1pass}), is
evaluated at the current parameter values. From our experience with
both implementations the one-pass one tends to require more iterations
to converge to the same solution, with the two implementations having
the same computational complexity per iteration. In addition, the
two-pass implementation requires starting values only for $\beta$,
while the one-pass implementation requires starting values for both
$\beta$ and $h_1, \ldots, h_n$.

\section{Demonstration: Diverted US flights in 2000}
\label{sec:air2000}

We demonstrate here the one-pass and two-pass implementations using
data on all $5\,683\,047$ commercial flights within the USA in 2000. The
data set is part of the data that was shared during The Data
Exposition Poster Session of the Graphics Section of the Joint
Statistical meetings in 2009, and is available at the Harvard
Dataverse data repository \citep{air2000:2009}. 

\begin{table}[t!]
\caption{Estimates and estimated standard errors (in parenthesis) for selected parameters of model~(\ref{eq:air_probit}) through ML, mBR and mJPL with one- and two-pass IWLS implementations. The table also reports the elapsed time, number of iterations, and average time per iteration for each method and implementation. The parameters corresponding to reference categories are set to $0$. The first and second column under the ML heading show the ML estimates when allowing for $15$ and $20$ IWLS iterations.}
\begin{center}
\begin{tabular}{llD{.}{.}{3}D{.}{.}{3}D{.}{.}{3}D{.}{.}{3}D{.}{.}{3}D{.}{.}{3}}
\toprule
  & &
\multicolumn{2}{c}{ML} & 
\multicolumn{2}{c}{mBR} & 
\multicolumn{2}{c}{mJPL} \\ 
\cmidrule{5-8}
  & &
\multicolumn{1}{c}{} &
\multicolumn{1}{c}{} & 
\multicolumn{1}{c}{one-pass} & 
\multicolumn{1}{c}{two-pass} & 
\multicolumn{1}{c}{one-pass} & 
\multicolumn{1}{c}{two-pass}\\
\midrule
  $\alpha$ & & -5.52 & -6.38 & -4.10 & -4.10 & -4.12 & -4.12\\
  & & (4.89) & (53.18) & (0.35) & (0.35) & (0.36) & (0.36)\\ 
\vdots \\  
$\delta_1$ & AA        & 2.49 & 3.36 & 1.08 & 1.08 & 1.09 & 1.09\\
& & (4.89) & (53.18) & (0.35) & (0.35) & (0.36) & (0.36)\\
$\delta_2$ & AQ & 0 & 0 & 0 & 0 & 0 & 0 \\
\\
$\delta_3$ & AS        & 2.59 & 3.45 & 1.17 & 1.17 & 1.18 & 1.18\\
                       & & (4.89) & (53.18) & (0.35) & (0.35) & (0.36) & (0.36)\\
$\delta_4$ & CO        & 2.33 & 3.19 & 0.91 & 0.91 & 0.92 & 0.92\\
                       & & (4.89) & (53.18) & (0.35) & (0.35) & (0.36) & (0.36)\\
$\delta_5$ & DL        & 2.36 & 3.22 & 0.94 & 0.94 & 0.95 & 0.95\\
                      &  & (4.89) & (53.18) & (0.35) & (0.35) & (0.36) & (0.36)\\
$\delta_6$ & HP        & 2.18 & 3.05 & 0.76 & 0.76 & 0.78 & 0.78\\
                      &  & (4.89) & (53.18) & (0.35) & (0.35) & (0.36) & (0.36)\\
$\delta_7$ & NW        & 2.44 & 3.30 & 1.02 & 1.02 & 1.04 & 1.04\\
                     &   & (4.89) & (53.18) & (0.35) & (0.35) & (0.36) & (0.36)\\
$\delta_8$ & TW        & 2.36 & 3.23 & 0.94 & 0.94 & 0.96 & 0.96\\
                      &  & (4.89) & (53.18) & (0.35) & (0.35) & (0.36) & (0.36)\\
$\delta_9$ & UA        & 2.35 & 3.22 & 0.94 & 0.94 & 0.95 & 0.95\\
                      &  & (4.89) & (53.18) & (0.35) & (0.35) & (0.36) & (0.36)\\
$\delta_{10}$ & US        & 2.54 & 3.41 & 1.12 & 1.12 & 1.14 & 1.14\\
                      &  & (4.89) & (53.18) & (0.35) & (0.35) & (0.36) & (0.36)\\
$\delta_{11}$ & WN        & 2.41 & 3.28 & 0.99 & 0.99 & 1.01 & 1.01\\
                      &  & (4.89) & (53.18) & (0.35) & (0.35) & (0.36) & (0.36)\\ 
  \vdots & \\
\midrule
\multicolumn{2}{l}{Time (sec)}  & 215.55 & 273.09 & 219.23 & 379.66 & 217.04 & 378.72 \\
\multicolumn{2}{l}{Iterations} & 15 & 20 & 12 & 12 & 12 & 12\\
\multicolumn{2}{l}{Time/Iteration (sec)} & 14.37 & 13.65 & 18.27 & 31.64 & 18.09 & 31.56 \\
\bottomrule
\end{tabular}
\end{center}
\label{tab:diverted_model}
\end{table}

Suppose that interest is in modelling the probability of diverted US
flights in 2000 in terms of the departure date attributes, the
scheduled departure and arrival times, the coordinates and distance
between departure and planned arrival airports, and the
carrier. Towards this goal, we assume that the diversion status of the
$i$th flight is a Bernoulli random variable with probability $\pi_i$ modelled
as
\begin{equation}
  \label{eq:air_probit}
  \Phi^{-1}(\pi_i) = \alpha + \beta^\top M_i + \gamma^\top W_i +   \delta^\top C_i + \zeta_{(d)}   T_{(d),i} + \zeta_{(a)} T_{(a),i}  + \rho D_i + \psi_{(d)}^\top L_{(d), i} + \psi_{(a)}^\top L_{(a), i} \, ,
\end{equation}
and that diversions are conditionally independent given the covariate
information that appears in~(\ref{eq:air_probit}). The covariate
information consists of $M_i$, which is a vector of $12$ dummy
variables characterizing the calendar month that the $i$th flight took
place; $W_i$, which is a vector of $7$ dummy variables characterizing
the week day that the $i$th flight took place; $C_i$, which is a
vector of $11$ dummy variables, characterizing the carrier of the
$i$th flight; $T_{(d),i}$ and $T_{(a),i}$, which are the planned
departure and arrival times in a 24 hour format, respectively; $D_i$,
which is the distance between the origin and planned destination
airport, respectively, and; $L_{(d), i}$ and $L_{(a), i}$, which are the
$(x, y, z)$ coordinates of the departure and arrival airport,
respectively, computed from longitude (lon) and latitude (lat)
information as $x = \cos(\rm lat) \cos(\rm lon)$,
$y = \cos(\rm lat) \sin(\rm lon)$, and $z = \sin(\rm lat)$.

For identifiability reasons, we fix $\beta_{1} = 0$ (January as a
reference month), $\gamma_1 = 0$ (Monday as a reference day)
and $\delta_2 = 0$ (carrier AQ as a reference carrier). Ignoring the
columns corresponding to those parameters, the model matrix for the
model with linear predictor as in~(\ref{eq:air_probit}) has dimension
$5683047 \times 37$, which requires about $1.6$GB of memory, which is,
nowadays, manageable by the available memory in many mid-range laptops.

The reasons we choose this data set for the demonstration of
bounded-memory fitting procedures are data availability (data is
publicly available with an permissive license), reproducibility (the
data set is manageable in average- to high-memory hardware
configurations, even when copying takes place, so that the majority of
readers can reproduce the numerical results using the scripts we
provide in the Supplementary Material), and occurrence of infinite
estimates in a data set that is orders of magnitude larger than the
data sets used in published work and software on data separation and
infinite ML estimates in binomial-response GLMs. For the latter,
testing for data separation by solving, for example, the
\citet{konis:2007} linear programs is computationally demanding. For
example, the \texttt{detect\_infinite\_estimates()} method of the {\tt
  detectseparation} R package did not complete even after 12 hours of
computation on a 2021 MacBook Pro with an Apple M1 Max chip and 64 GB
RAM.

Table~\ref{tab:diverted_model} shows the ML estimates of the
parameters $\alpha$ and $\delta$ of model~\eqref{eq:air_probit} after
15 and 20 IWLS iterations using the bounded-memory procedures in
Section~\ref{sec:bmem_al} as implemented in the \texttt{biglm} R
package \citep{lumley:2020}, using chunks of $c = 10\, 000$
observations, $\epsilon = 10^{-3}$ for the convergence criteria in
Section~\ref{sec:iwls_adj}, and setting $\beta^{(0)}$ to a vector of
$37$ zeros. The estimates and estimated standard errors for $\alpha$
and the components of $\delta$ grow in absolute value with the number
of IWLS iterations, which is typically the case when the maximum
likelihood estimates are infinite \citep[see, for example,][for such
behaviours in the case of multinomial logistic
regression]{lesaffre+albert:1989}.  The ML estimates for the other
parameters do not change in the reported accuracy as we move from 15
to 20 IWLS iterations; see Table~\ref{tab:diverted_model_all} of the
Supplementary Materials document for estimates for all parameters. In
contrast, mBR and mJPL return finite estimates for all parameters by
declaring convergence before reaching the limit of allowable
iterations, for both their one- and two-pass implementations. As
expected by the discussion in Section~\ref{sec:mem_cpu}, the one-pass
implementations require about $58\%$ of the time per iteration that
the two-pass implementations require. The fact that, for this
particular data set, the one- and two-pass implementations of mBR and
mJPL required the same number of iterations is a coincidence and is
not necessary the case for other data sets. No memory issues have been
encountered obtaining the ML, mJPL and mBR fits, even when the fits
were re-computed on Ubuntu virtual machines with 2 cores and 2GB and
4GB of RAM, where also no swapping took place. On the other hand, the
mBR and mJPL fits could not be obtained in those virtual environments
using the \texttt{brglm2} R package, because available physical memory
was exhausted.

Model~(\ref{eq:air_probit}) has also been fit using the
\texttt{brglm2} R package on a 2021 MacBook Pro with an Apple M1 Max
chip and 64 GB RAM, which can comfortably handle having copies of the
whole data set in memory, using the same starting values and
convergence criterion, and ensuring that \texttt{brglm2} carries out
the IWLS update~(\ref{eq:iwlsadj}) with no modifications. The mBR and
mJPL fits required $313.29$ and $315.87$ seconds, respectively, to
complete in $12$ iterations. The estimates from \texttt{brglm2} are
the same to those shown in Table~\ref{tab:diverted_model} (and
Table~\ref{tab:diverted_model_all} of the Supplementary Materials
document), and, hence, are not reported. This is as expected, because
as discussed in Section~\ref{sec:two-pass}, the one- and two-pass
implementation have the correct stationary point.

The observations made in this example highlight that, even for large
data sets, use of estimation methods that return estimates in the
interior of the parameter space is desirable, especially so when the
computational cost for an IWLS iteration for ML scales linearly with
the number of observations.

\section{Privacy-preserving estimation}
\label{sec:privacy}

As noted by a reviewer, IWLS in chunks
(Section~\ref{sec:chunkwiseIWLS} and
Algorithm~\ref{algo:supp_incrIWLS} in the Supplementary Material
document) for ML estimation, and the one-pass implementation for
adjusted score estimation in chunks (Section~\ref{sec:one-pass}, and
Algorithm~\ref{algo:supp_incrIWLS1} in the Supplementary Material
document) may be readily adapted to fit GLMs when distinct parts of
the data is stored across different sites and, due to privacy
concerns, cannot be fully transferred across sites. Such adaptations
provide guarantees of stability of all numerical and inferential
procedures even when the data set is separated and infinite ML
estimates occur, in settings where checking for infinite estimates is
not feasible with existing algorithms.

Suppose there are $K$ sites, with the $k$ site holding data
$\{X_{:k}, y_{:k}\}$, that $\phi = 1$ (e.g. for binomial and Poisson
responses), and estimation is by maximum likelihood. First, the
current value $\beta^\dagger$ ($p$ real numbers) for the estimates is
broadcast to all sites. Site 1 computes the weights $w_{:1}$ and the
working variates $z_{:1}$ for the data $\{X_{:1}, y_{:1}\}$ it holds
at $\beta^\dagger$, and uses those to compute $\bar{R}_{:1}$ and
$\bar{b}_{:1}$, which are then transmitted to Site 2 ($p(p + 3)/2$
real numbers). Site 2 computes the weights $w_{:2}$ and the working
variates $z_{:2}$ for the data $\{X_{:2}, y_{:2}\}$ it holds at
$\beta^\dagger$, and uses those to update $\bar{R}_{:1}$ and
$\bar{b}_{:1}$ to $\bar{R}_{:2}$ and $\bar{b}_{:2}$, which are then
communicated to Site 3 ($p(p + 3)/2$ real numbers), and so on. Once
all $K$ sites have been visited, a compute node (which may be one of
the sites) takes $\bar{R} = \bar{R}_{:K}$ and
$\bar{b} = \bar{b}_{:K}$, computes $\beta^* = \bar{R}^{-1} \bar{b}$,
and checks if $\| \beta^* - \beta^\dagger \| < \epsilon$, for some
$\epsilon$. If that holds, then the process ends and the estimates
$\hat\beta := \beta^*$ are returned along with $\bar{R}$, which can be
used for the computation of standard errors. Otherwise,
$\beta^\dagger$ is set to $\beta^*$ and the sites are visited again.

\begin{figure}[t]
  \caption{A flowchart that displays the adaptation of incremental
    IWLS (Algorithm~\ref{algo:supp_incrIWLS}) for the estimation of
    GLMs when distinct parts of the data is stored across three
    different sites and, due to privacy concerns, cannot be fully
    transferred across sites.}
  \begin{center}
    \includegraphics[height = 0.59\textheight]{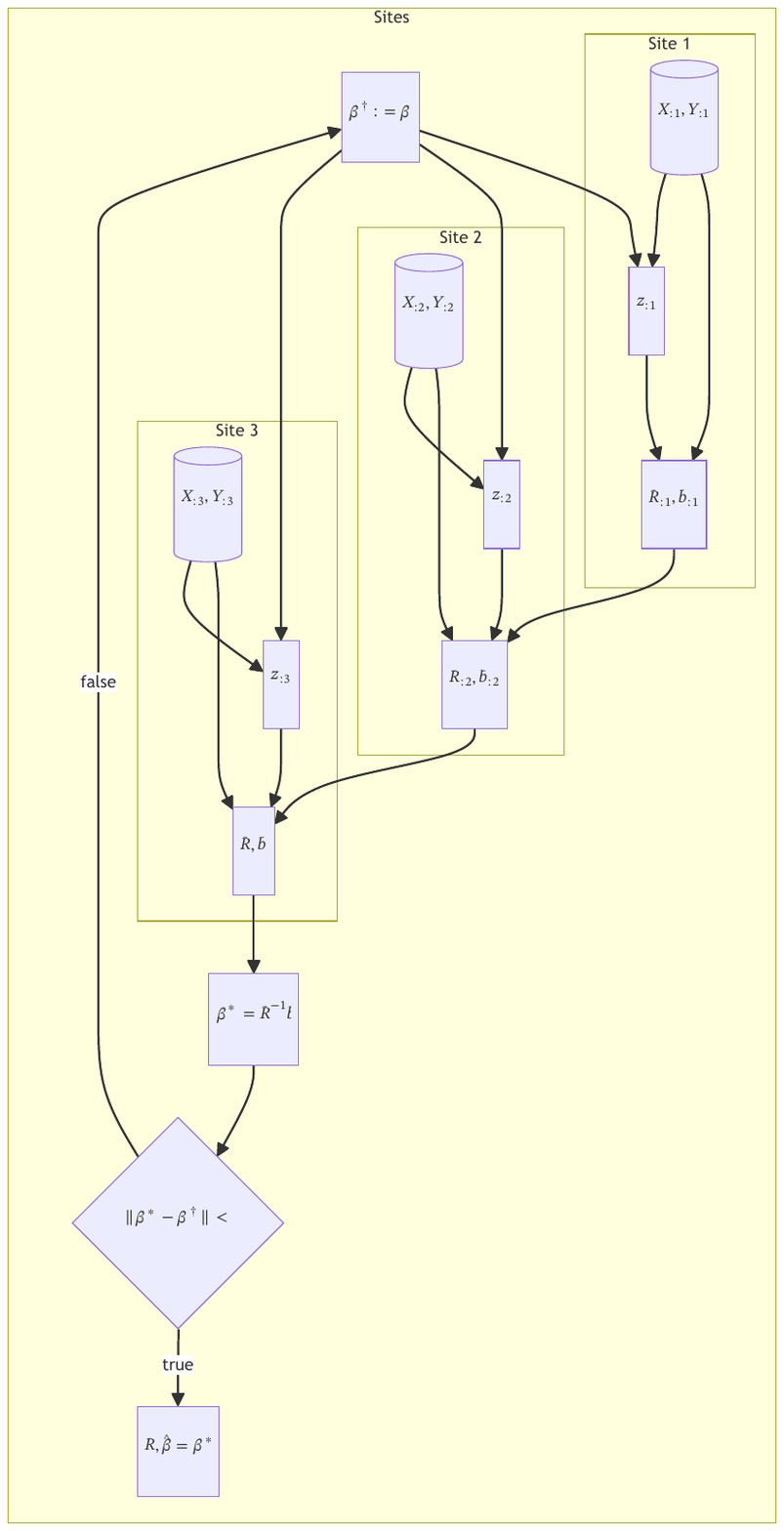}
  \end{center}
  \label{fig:sites}
\end{figure}

Figure~\ref{fig:sites} displays the process for $K = 3$. For unknown
$\phi$, at each iteration, each site updates the current value of the
sum of squared residuals $\sum w_i (z_i - \eta_i)$ and passes that to
the next site along with $\bar{R}_{:k}$ and $\bar{b}_{:k}$. Once all
sites have been visited, $\phi$ is updated as
in~(\ref{eq:phi_mom_update}); see line~\ref{incrIWLS-SSR} of
Algorithm~\ref{algo:supp_incrIWLS}.

For the one-pass implementation of mBR and mJPL, the sites should
retain two consecutive values of the estimates, the one that was
broadcast in the previous iteration and the one that has been
broadcast in the current iteration. After the first visit to all
sites, the value of $\bar{R}^{-1}$ is also broadcast to the sites
along with $\beta^\dagger$ ($p(p + 3)/2$ real numbers in total). Then, each
site updates $z$ according to lines
\ref{incrIWLS1-start}-\ref{incrIWLS1-end} in
Algorithm~\ref{algo:supp_incrIWLS1}.

The two-pass implementation of mBR and mJPL can also be adapted when
the data is stored across different sites, with only a bit of added
complexity in design. Each site should be visited twice per iteration,
and the sites will perform different computations in the first and the
second visit (see Algorithm~\ref{algo:supp_incrIWLS2} in the
Supplementary Material Document).

\section{Concluding remarks}
\label{sec:concluding}

We have developed two variants of IWLS that can estimate the
parameters of GLMs using adjusted score equations for mean bias
reduction and maximum Jeffreys'-prior penalized likelihood, for data
sets that exceed computer memory or even hard-drive capacity and are
stored in remote databases. The two procedures numerically solve the
mBR and mJPL adjusted score equations, exactly as the in-memory
methods of \citet{kosmidis+kennepagui+sartori:2020} do, and the
estimates they return are invariant to the choice of $c$ or the
ordering of the data chunks. We used both procedures in
Section~\ref{sec:air2000} to obtain finite estimates of the parameters
of a probit regression model with $37$ parameters from about $5.5$
million observations, where the ML estimates have been found to have
infinite components.

The choice of the chunk-size $c$ should be based on how much memory
the practitioner wants to use or has available. Choosing a large value
of $c$, or equivalently, keeping a small value of $K$ (the number of
chunks) is beneficial, especially if operations can be vectorized. It
is difficult to objectively quantify the impact of the choice of $c$
in execution time because that impact depends in all implementation,
computing framework, speed of data retrieval, and available hardware.

As \citet{kosmidis+kennepagui+sartori:2020} show, median bias
reduction for $\beta$ can also be achieved using an IWLS procedure
after modifying the ML working variates. The IWLS update for median
bias reduction has the form
\[
\beta^{(j+1)} := \left(X^\top W^{(j)} X\right)^{-1} X^\top W^{(j)}\left(z^{(j)} + \phi^{(j)} \left\{ H^{(j)} \xi^{(j)} + X u^{(j)} \right\}\right)\, .
\]
The particular form of the $p$-vector $u$ is given in
\citet[expression~(10)]{kosmidis+kennepagui+sartori:2020}, and depends
on the inverse of $i_{\beta\beta}$ and hence on $\bar{R}$. Since $X u$
can be computed in a chunkwise manner for any given $u$, it is
possible to develop one- and two-pass implementations of the IWLS
procedure for median bias reduction by the same arguments as those
used in Section~\ref{sec:two-pass} and
Section~\ref{sec:one-pass}. These procedures are computationally more
expensive than the procedures for mBR and mJPL because each component
of $u$ requires $O(np^3)$ operations.

Both the one-pass and two-pass IWLS implementations have the correct
stationary point, that is the solution of the adjusted score
equations~(\ref{eq:adj_eq_beta}) for any GLM, as
that would be obtained having the whole data set in memory. We should
note though that a formal account of their convergence properties is
challenging for all GLMs and any combination of
adjustments in the adjusted score equations~(\ref{eq:adj_eq_beta})
and~(\ref{eq:adj_eq_phi}). The two-pass implementation, in particular,
is formally equivalent to carrying out the IWLS
update~(\ref{eq:iwlsadj}) with the full data in memory, which is what
popular and well-used in practice and simulation experiments software
like \texttt{brglm2} and \texttt{logistf} R packages implement. In all
our numerical experiments we did not encounter any cases where the
two-pass implementation did not converge. Some progress with
convergence analysis of the IWLS update~(\ref{eq:iwlsadj}) may be
possible by noting that the IWLS update for the adjusted score
equations is formally a quasi-Newton iteration, where only the leading
term of the Jacobian of the adjusted score functions is used in the
step calculation. In particular, the Jacobian of the adjusted score
functions can be written as $i + \delta + A$, where $i$ is the
expected information matrix~(\ref{eq:information}) and hence $O(n)$,
$\delta = O_p(n^{1/2})$ with $\expect(\delta) = 0_{p \times p}$, and
$A = O(1)$ is the Jacobian of the score adjustments.

The one-pass implementation, despite of having a faster iteration than
the two-pass one, it tends to require more iterations to converge, and
is more sensitive on starting values. It may be worthwhile to consider
combinations of the two implementations, where one starts with the two
pass and switches to the one-pass once the difference between
consecutive parameter values is small enough in some appropriate
sense.

Current work focuses on reducing the cubic complexity on $p$, without
impacting the finiteness and bias reducing properties of
mJPL. \citet{drineas+etal:2012} on the approximation of the leverage
seems relevant.

\section{Supplementary materials}

The Supplementary Material provides the Supplementary Material
document that is cross-referenced above and contains an exposition of
Givens rotations (Section~\ref{sec:supp_givens}), pseudo-code for
iteratively reweighted least squares in chunks
(Algorithm~\ref{algo:supp_incrIWLS}), pseudo-code for the one- and
two-pass implementations for solving the adjusted scores equations in
chunks (Algorithm~\ref{algo:supp_incrIWLS1} and
Algorithm~\ref{algo:supp_incrIWLS1}, respectively), and all numerical
results from the case study of Section~\ref{sec:air2000}. The
Supplementary Material also provides R code to reproduce all numerical
results in the main text and in the Supplementary Material
document. The R code is organized in the two directories {\tt
  diverted-flights} and {\tt biglm}. The former directory provides
code for the case study of Section~\ref{sec:air2000}; the
\texttt{README} file provides specific instructions to reproduce the
results, along with the versions of the contributed R packages that
have been used to produce the results in the main text. The {\tt
  biglm} directory has a port of the \texttt{biglm} R package
\citep{lumley:2020}, which implements the one- and two-pass IWLS
variants for solving the bias-reducing adjusted score equations
\citep{firth:1993} and for maximum Jeffreys'-penalized likelihood
estimation \citep{kosmidis+firth:2021}. The Supplementary Material is
available at
\url{https://github.com/ikosmidis/bigbr-supplementary-material}.

\section{Declarations}

For the purpose of open access, the authors have applied a Creative
Commons Attribution (CC BY) licence to any Author Accepted Manuscript
version arising from this submission.

\bibliographystyle{jss2}
\bibliography{bigbr}



\section*{Supplementary material}

defined as

$$G(i,j,\theta) = \begin{array}{c} \\ \\ i \\ \\ j \\ \\ \\ \end{array} \begin{bmatrix} 1 & \cdots & 0 & \cdots & 0 & \cdots & 0 \\ \vdots & \ddots & \vdots & & \vdots & & \vdots \\ 0 & \cdots & c_\theta & \cdots & s_\theta & \cdots & 0 \\ \vdots & & \vdots & \ddots & \vdots & & \vdots \\ 0 & \cdots & -s_\theta & \cdots & c_\theta & \cdots & 0 \\ \vdots & & \vdots & & \vdots & \ddots & \vdots \\ 0 & \cdots & 0 & \cdots & 0 & \cdots & 1 \end{bmatrix} \begin{array}{c} i \qquad\qquad j \\ \\ \\ \\ \\ \\ \\ \end{array}$$

where $c_\theta = \cos\theta$ and $s_\theta = \sin\theta$. Then, $G(i,j,\theta)^\top r$ is a counter-clockwise rotation of $\theta$ radians of $r$ in the $(i,j)$ coordinate plane, $G(i,j,\theta)$ is orthogonal ($G(i,j,\theta)^\top = G(i,j,\theta)^{-1}$), and the product $G(i,j,\theta)^\top R$ affects just the $i$, $j$ rows of $R$. We can use Givens rotations to eliminate arbitrary elements of $R$. For example, we can eliminate the $(3,2)$ element of $R$ by taking the product $G(2,3,\theta)^\top R$ and choosing $s_\theta$ and $c_\theta$ as the solution to the system of two equations $s_\theta^2 + c_\theta^2 = 1$ and $s_\theta r_{22} + c_\theta r_{32} = 0$ with respect to $s_\theta$ and $c_\theta$; see, for example, Golub and Van Loan (2013, Algorithm 5.1.3) for an algorithm that solves that system. Note that there is no need to compute $\theta$.

## S2 Procedures `updateQR`, `incrLS`, `incrIWLS`, `incrIWLS1`, `incrIWLS2`

---

**Algorithm S1** Update the current value of $\bar{R}$ and $\bar{b}$ (`Rbar` and `bbar`, respectively) for the regression of $b$ on $A$ with observations weights $w$, given chunks of $A$, $b$, and $w$ (`Ac`, `bc`, and `wc`, respectively). `INCL` is a streamlined interface to the `INCLUD` implementation in Miller (1992, Algorithm AS274.1). Specifically, `INCL(p, Ai, bi, wi, Rbar, bbar)` returns the updated values of `Rbar` and `bbar` given a row `Ai` of `Ac`, an element `bi` of `bc`, and an element `wi` of `wc`. `INCL` internally calls `INCLUD(p, p(p − 1)/2, wi, Ai, bi, D̃′, R̃′, b̃, 0, 0)`, which updates $\tilde{R}$ (an upper triangular matrix with ones or zeros in the diagonal), $\tilde{D}$ (a diagonal matrix), and $\tilde{b}$, where $\tilde{R}'$ is a vector with the upper-triangular elements of $\tilde{R}$ stored by row and excluding the diagonal elements, and $\tilde{D}'$ is the diagonal elements of $\tilde{D}$. In the notation of the current work, $\tilde{D}^{1/2}\tilde{R} = \bar{R}$, $\tilde{D}^{1/2}\tilde{b} = \bar{b}$.

1: **procedure** UPDATEQR(c, p, Ac, bc, wc, Rbar, bbar)
2:     **for** i $\in \{1, 2, \ldots, $ c$\}$ **do**
3:         Ai $\leftarrow$ [Ac[i, 1], ..., Ac[i, p]]
4:         bi $\leftarrow$ bc[i]
5:         wi $\leftarrow$ wc[i]
6:         {Rbar, bbar} $\leftarrow$ INCL(p, Ai, bi, wi, Rbar, bbar)
7:     **end for**
8:     **return** {Rbar, bbar}
9: **end procedure**

---

**Algorithm S2** `incrLS` takes as input (a storage location for) $A$, $b$, and $w$, and a chunk size $c$. Internally, `updateQR` is used to compute the least squares estimate $\hat{\psi}$ (`psihat`) of the regression of $b$ on $A$ with observation weights $w$, by applying it sequentially on all chunks of $c$ observations, with the last chunk containing $c$ or less observations.

```
 1: procedure INCRLS(A, b, c, w)
 2:     n ← numberofrows(A)          ▷ must be equal to length(b) and length(w), and greater or equal to c
 3:     p ← numberofcolumns(A)
 4:     Rbar ← a p × p matrix of zeros
 5:     bbar ← a vector of p zeros
 6:     K ← ⌈n/c⌉                    ▷ ⌈a⌉ is the ceiling of a
 7:     for k ∈ {1, 2, ..., K} do
 8:         i1 ← (k − 1) ∗ c + 1
 9:         if k < K then
10:             i2 ← k ∗ c
11:         else
12:             c ← n − i1 + 1
13:             i2 ← n
14:         end if
15:         Ac ← A[i1 : i2, 1 : p]
16:         bc ← b[i1 : i2]
17:         wc ← w[i1 : i2]
18:         {Rbar, bbar} ← updateQR(c, p, Ac, bc, wc, Rbar, bbar)
19:     end for
20:     psihat ← inverse(Rbar) · bbar    ▷ using back substitution; · is matrix product
21:     return {psihat, Rbar}
22: end procedure
```

**Algorithm S3** `incrIWLS` is an implementation of incremental IWLS for generalized linear models. It takes as input (a storage location for) $X$, $y$, and the observation weights $m$, the chunk size $c$, an implementation of the inverse link function $F$ and its derivative $f$, an implementation of the variance function $V$, starting values for $\hat{\beta}$ (`betahat`), a tolerance $\epsilon$ (`epsilon`), and the maximum allowed number of IWLS iterations (`maxit`). `incrIWLS` sequentially goes through all chunks of $X$ and $y$, updates the corresponding chunk of the working variates $z$ at the current value of $\hat{\beta}$, and uses `updateQR` in Algorithm S1 to update the $\bar{R}$ and $\bar{b}$ from the chunks of $z$, $X$, and $w$. Once all chunks have been visited the weighted least squares estimate $\hat{\beta}$ is computed. The process is repeated until the norm of the difference of the current iterate of $\hat{\beta}$ from the previous one is smaller than $\epsilon$, or the maximum number of allowed iterations is exceeded.

```
 1: procedure INCRIWLS(X, y, m, c, F, f, V, betahat, epsilon, maxit)
 2:     n ← numberofrows(X)         ▷ must be equal to length(y) and length(m), and greater or equal to c
 3:     p ← numberofcolumns(X)      ▷ must be equal to length(betahat) and greater or equal to c
 4:     df ← n − p
 5:     K ← ⌈n/c⌉                   ▷ ⌈a⌉ is the ceiling of a
 6:     iter ← 1
 7:     Rbar ← a p × p matrix of zeros
 8:     bbar ← a vector of p zeros
 9:     SSR ← 0
10:     phihat ← 1                  ▷ if GLM has an unknown dispersion parameter
11:     for k ∈ {1, 2, ..., K} do
12:         i1 ← (k − 1) ∗ c + 1
13:         if k < K then
14:             i2 ← k ∗ c
15:         else
16:             c ← n − i1 + 1
17:             i2 ← n
18:         end if
19:         Xc ← X[i1 : i2, 1 : p]
20:         yc ← y[i1 : i2]
21:         mc ← m[i1 : i2]
22:         etac ← Xc · betahat                       ▷ · is matrix product
23:         muc ← [F(etac[1]), ..., F(etac[c])]
24:         dc ← [f(etac[1]), ..., f(etac[c])]
25:         vc ← [V(muc[1]), ..., V(muc[c])]
26:         wc ← mc ∗ dc ∗ dc/vc                      ▷ elementwise ∗, /
27:         zc ← etac + (yc − etac)/dc                ▷ elementwise +, −, /
28:         SSR ← SSR + sum(wc ∗ (zc − eta)²)         ▷ if GLM has an unknown dispersion parameter
29:         {Rbar, bbar} ← updateQR(c, p, Xc, zc, wc, Rbar, bbar)
30:     end for
31:     beta0 ← betahat                               ▷ if GLM has an unknown dispersion parameter
32:     RbarInv ← inverse(Rbar)                       ▷ using back substitution
33:     betahat ← RbarInv · bbar
34:     phihat ← SSR/df
35:     if ||betahat − beta0|| > epsilon then
36:         if iter < maxit then
37:             iter ← iter + 1
38:             Go to 7
39:         else
40:             print algorithm not converged
41:             return {betahat, Rbar}
42:         end if
43:     else
44:         return {betahat, phihat, Rbar}
45:     end if
46: end procedure
```

**Algorithm S4** incrIWLS1 is an implementation of the one-pass incremental IWLS for the adjusted score estimation of generalized linear models. It takes as input (a storage location for) $X$, $y$, and the observation weights $m$, the chunk size $c$, an implementation of the inverse link function $F$ and its first and second derivative $f$ and $f'$ (fdash), respectively, an implementation of the variance function $V$ and its derivative $V'$ (Vdash), starting values for $\hat{\beta}$ (betahat), a tolerance $\epsilon$ (epsilon), and the maximum allowed number of IWLS iterations (maxit). The mJPL estimates are computed if mjpl is true. Otherwise, the mBR estimates are computed.

1: **procedure** INCRIWLS1(X, y, m, c, F, f, fdash, V, Vdash, betahat, epsilon, maxit, mjpl)
2:     Lines 2-28 from Algorithm S3
3:     **if** iter $> 1$ **then**
4:         etac0 $\leftarrow$ Xc $\cdot$ beta0                                                   ▷ $\cdot$ is matrix product
5:         muc0 $\leftarrow$ [F(etac0[1]), ..., F(etac0[c])]
6:         dc0 $\leftarrow$ [f(etac0[1]), ..., f(etac0[c])]
7:         vc0 $\leftarrow$ [V(muc0[1]), ..., V(muc0[c])]
8:         wc0 $\leftarrow$ mc $*$ dc0 $*$ dc0/vc0                                         ▷ elementwise $*$, $/$
9:         xR0 $\leftarrow$ Xc $\cdot$ RbarInv
10:         hc0 $\leftarrow$ (xR0 $\cdot$ t(xR0)) $*$ wc0                ▷ t(A) is the transpose of matrix A; elementwise $*$
11:         ddashc $\leftarrow$ [fdash(etac[1]), ..., fdash(etac[c])]
12:         hkappac $\leftarrow$ hc0 $*$ ddashc/(2 $*$ dc $*$ wc)                         ▷ elementwise $*$, $/$
13:         zc $\leftarrow$ zc $+$ phihat $*$ hkappac                                         ▷ elementwise $+$
14:         **if** mjpl $=$ true **then**
15:             vdashc $\leftarrow$ [Vdash(muc[1]), ..., Vdash(muc[c])]
16:             zc $\leftarrow$ zc $+$ phihat $*$ (hkappac $-$ hc0 $*$ vdashc/(2 $*$ mc $*$ dc))     ▷ elementwise $+$, $-$, $*$, $/$
17:         **end if**
18:     **end if**
19:     Lines 29-45 from Algorithm S3
20: **end procedure**

**Algorithm S5** incrIWLS2 is an implementation of the two-pass incremental IWLS for the adjusted score estimation of generalized linear models. It takes as input (a storage location for) $X$, $y$, and the observation weights $m$, the chunk size $c$, an implementation of the inverse link function $F$ and its first and second derivative $f$ and $f'$ (fdash), respectively, an implementation of the variance function $V$ and its derivative $V'$ (Vdash), starting values for $\hat{\beta}$ (betahat), a tolerance $\epsilon$ (epsilon), and the maximum allowed number of IWLS iterations (maxit). The mJPL estimates are computed if mjpl is true. Otherwise, the mBR estimates are computed.

1: **procedure** INCRIWLS2(X, y, m, c, F, f, fdash, V, Vdash, betahat, epsilon, maxit, mjpl)
2:     Lines 2-32 from Algorithm S3
3:     **for** $k \in \{1, 2, \ldots, K\}$ **do**     ▷ Second pass
4:         i1 ← (k − 1) ∗ c + 1
5:         **if** k < K **then**
6:             i2 ← k ∗ c
7:         **else**
8:             c ← n − i1 + 1
9:             i2 ← n
10:         **end if**
11:         Xc ← X[i1 : i2, 1 : p]
12:         mc ← m[i1 : i2]
13:         etac ← Xc · betahat     ▷ · is matrix product
14:         muc ← [F(etac[1]), . . . , F(etac[c])]
15:         dc ← [f(etac[1]), . . . , f(etac[c])]
16:         vc ← [V(muc[1]), . . . , V(muc[c])]
17:         wc ← mc ∗ dc ∗ dc/vc     ▷ elementwise ∗, /
18:         xR ← Xc · RbarInv
19:         hc ← (xR · t(xR)) ∗ wc     ▷ t(A) is the transpose of matrix A; elementwise ∗
20:         ddashc ← [fdash(etac[1]), . . . , fdash(etac[c])]
21:         hkappac ← phihat ∗ hc ∗ ddashc/(2 ∗ dc ∗ wc)     ▷ elementwise ∗, /
22:         **if** mjpl = true **then**
23:             vdashc ← [Vdash(muc[1]), . . . , Vdash(muc[c])]
24:             hkappac ← 2 ∗ hkappac − phihat ∗ hc ∗ vdashc/(2 ∗ mc ∗ dc)     ▷ elementwise ∗, −, +
25:         **end if**
26:         {Rbar, bbarA} ← updateQR(c, p, Xc, hkappac, wc, Rbar, bbarA)
27:     **end for**
28:     betahat ← RbarInv · (bbar + bbarA)     ▷ elementwise +
29:     Lines 34-45 from Algorithm S3
30: **end procedure**

# S3 Estimates and estimated standard errors for the parameters of model (14)

Table S1: Estimates and estimated standard errors for the parameters of model (14) through ML, mBR and mJPL with one- and two-pass IWLS implementations. The table also reports the elapsed time, number of iterations, and average time per iteration for each method and implementation. The parameters corresponding to reference categories are set to 0. The first and second column under the ML heading show the ML estimates when allowing for 15 and 20 IWLS iterations.

|            |           | ML        |           | mBR       |           | mJPL      |           |
|            |           |           |           | 1-pass    | 2-pass    | 1-pass    | 2-pass    |
|------------|-----------|-----------|-----------|-----------|-----------|-----------|-----------|
| $\alpha$   |           | $-5.52$   | $-6.38$   | $-4.10$   | $-4.10$   | $-4.12$   | $-4.12$   |
|            |           | $(4.89)$  | $(53.18)$ | $(0.35)$  | $(0.35)$  | $(0.36)$  | $(0.36)$  |
| $\beta_1$  | January   | 0         | 0         | 0         | 0         | 0         | 0         |
| $\beta_2$  | February  | $-0.04$   | $-0.04$   | $-0.04$   | $-0.04$   | $-0.04$   | $-0.04$   |
|            |           | $(0.01)$  | $(0.01)$  | $(0.01)$  | $(0.01)$  | $(0.01)$  | $(0.01)$  |
| $\beta_3$  | March     | $-0.10$   | $-0.10$   | $-0.10$   | $-0.10$   | $-0.10$   | $-0.10$   |
|            |           | $(0.01)$  | $(0.01)$  | $(0.01)$  | $(0.01)$  | $(0.01)$  | $(0.01)$  |
| $\beta_4$  | April     | $-0.08$   | $-0.08$   | $-0.08$   | $-0.08$   | $-0.08$   | $-0.08$   |
|            |           | $(0.01)$  | $(0.01)$  | $(0.01)$  | $(0.01)$  | $(0.01)$  | $(0.01)$  |
| $\beta_5$  | May       | $0.01$    | $0.01$    | $0.01$    | $0.01$    | $0.01$    | $0.01$    |
|            |           | $(0.01)$  | $(0.01)$  | $(0.01)$  | $(0.01)$  | $(0.01)$  | $(0.01)$  |
| $\beta_6$  | June      | $0.06$    | $0.06$    | $0.06$    | $0.06$    | $0.06$    | $0.06$    |
|            |           | $(0.01)$  | $(0.01)$  | $(0.01)$  | $(0.01)$  | $(0.01)$  | $(0.01)$  |
| $\beta_7$  | July      | $-0.01$   | $-0.01$   | $-0.01$   | $-0.01$   | $-0.01$   | $-0.01$   |
|            |           | $(0.01)$  | $(0.01)$  | $(0.01)$  | $(0.01)$  | $(0.01)$  | $(0.01)$  |
| $\beta_8$  | August    | $-0.00$   | $-0.00$   | $-0.00$   | $-0.00$   | $-0.00$   | $-0.00$   |
|            |           | $(0.01)$  | $(0.01)$  | $(0.01)$  | $(0.01)$  | $(0.01)$  | $(0.01)$  |
| $\beta_9$  | September | $-0.09$   | $-0.09$   | $-0.09$   | $-0.09$   | $-0.09$   | $-0.09$   |
|            |           | $(0.01)$  | $(0.01)$  | $(0.01)$  | $(0.01)$  | $(0.01)$  | $(0.01)$  |
| $\beta_{10}$ | October | $-0.13$   | $-0.13$   | $-0.13$   | $-0.13$   | $-0.13$   | $-0.13$   |
|            |           | $(0.01)$  | $(0.01)$  | $(0.01)$  | $(0.01)$  | $(0.01)$  | $(0.01)$  |
| $\beta_{11}$ | November | $-0.15$  | $-0.15$   | $-0.15$   | $-0.15$   | $-0.15$   | $-0.15$   |
|            |           | $(0.01)$  | $(0.01)$  | $(0.01)$  | $(0.01)$  | $(0.01)$  | $(0.01)$  |
| $\beta_{12}$ | December | $0.03$   | $0.03$    | $0.03$    | $0.03$    | $0.03$    | $0.03$    |
|            |           | $(0.01)$  | $(0.01)$  | $(0.01)$  | $(0.01)$  | $(0.01)$  | $(0.01)$  |
| $\gamma_1$ | Monday    | 0         | 0         | 0         | 0         | 0         | 0         |
| $\gamma_2$ | Tuesday   | $-0.01$   | $-0.01$   | $-0.01$   | $-0.01$   | $-0.01$   | $-0.01$   |
|            |           | $(0.01)$  | $(0.01)$  | $(0.01)$  | $(0.01)$  | $(0.01)$  | $(0.01)$  |
| $\gamma_3$ | Wednesday | $-0.02$   | $-0.02$   | $-0.02$   | $-0.02$   | $-0.02$   | $-0.02$   |
|            |           | $(0.01)$  | $(0.01)$  | $(0.01)$  | $(0.01)$  | $(0.01)$  | $(0.01)$  |
| $\gamma_4$ | Thursday  | $0.05$    | $0.05$    | $0.05$    | $0.05$    | $0.05$    | $0.05$    |
|            |           | $(0.01)$  | $(0.01)$  | $(0.01)$  | $(0.01)$  | $(0.01)$  | $(0.01)$  |
| $\gamma_5$ | Friday    | $0.09$    | $0.09$    | $0.09$    | $0.09$    | $0.09$    | $0.09$    |
|            |           | $(0.01)$  | $(0.01)$  | $(0.01)$  | $(0.01)$  | $(0.01)$  | $(0.01)$  |
| $\gamma_6$ | Saturday  | $-0.02$   | $-0.02$   | $-0.02$   | $-0.02$   | $-0.02$   | $-0.02$   |

|  |  |  |  |  |  |  |  |
|---|---|---:|---:|---:|---:|---:|---:|
|  |  | (0.01) | (0.01) | (0.01) | (0.01) | (0.01) | (0.01) |
| $\gamma_7$ | Sunday | 0.07 | 0.07 | 0.07 | 0.07 | 0.07 | 0.07 |
|  |  | (0.01) | (0.01) | (0.01) | (0.01) | (0.01) | (0.01) |
| $\delta_1$ | AA | 2.49 | 3.36 | 1.08 | 1.08 | 1.09 | 1.09 |
|  |  | (4.89) | (53.18) | (0.35) | (0.35) | (0.36) | (0.36) |
| $\delta_2$ | AQ | 0 | 0 | 0 | 0 | 0 | 0 |
| $\delta_3$ | AS | 2.59 | 3.45 | 1.17 | 1.17 | 1.18 | 1.18 |
|  |  | (4.89) | (53.18) | (0.35) | (0.35) | (0.36) | (0.36) |
| $\delta_4$ | CO | 2.33 | 3.19 | 0.91 | 0.91 | 0.92 | 0.92 |
|  |  | (4.89) | (53.18) | (0.35) | (0.35) | (0.36) | (0.36) |
| $\delta_5$ | DL | 2.36 | 3.22 | 0.94 | 0.94 | 0.95 | 0.95 |
|  |  | (4.89) | (53.18) | (0.35) | (0.35) | (0.36) | (0.36) |
| $\delta_6$ | HP | 2.18 | 3.05 | 0.76 | 0.76 | 0.78 | 0.78 |
|  |  | (4.89) | (53.18) | (0.35) | (0.35) | (0.36) | (0.36) |
| $\delta_7$ | NW | 2.44 | 3.30 | 1.02 | 1.02 | 1.04 | 1.04 |
|  |  | (4.89) | (53.18) | (0.35) | (0.35) | (0.36) | (0.36) |
| $\delta_8$ | TW | 2.36 | 3.23 | 0.94 | 0.94 | 0.96 | 0.96 |
|  |  | (4.89) | (53.18) | (0.35) | (0.35) | (0.36) | (0.36) |
| $\delta_9$ | UA | 2.35 | 3.22 | 0.94 | 0.94 | 0.95 | 0.95 |
|  |  | (4.89) | (53.18) | (0.35) | (0.35) | (0.36) | (0.36) |
| $\delta_{10}$ | US | 2.54 | 3.41 | 1.12 | 1.12 | 1.14 | 1.14 |
|  |  | (4.89) | (53.18) | (0.35) | (0.35) | (0.36) | (0.36) |
| $\delta_{11}$ | WN | 2.41 | 3.28 | 0.99 | 0.99 | 1.01 | 1.01 |
|  |  | (4.89) | (53.18) | (0.35) | (0.35) | (0.36) | (0.36) |
| $\zeta_{(d)}$ | Departure time | 0.00 | 0.00 | 0.00 | 0.00 | 0.00 | 0.00 |
|  |  | (0.00) | (0.00) | (0.00) | (0.00) | (0.00) | (0.00) |
| $\zeta_{(a)}$ | Arrival Time | 0.00 | 0.00 | 0.00 | 0.00 | 0.00 | 0.00 |
|  |  | (0.00) | (0.00) | (0.00) | (0.00) | (0.00) | (0.00) |
| $\rho$ | Distance | 0.00 | 0.00 | 0.00 | 0.00 | 0.00 | 0.00 |
|  |  | (0.00) | (0.00) | (0.00) | (0.00) | (0.00) | (0.00) |
| $\psi_{(d),1}$ | Origin x | 0.03 | 0.03 | 0.03 | 0.03 | 0.03 | 0.03 |
|  |  | (0.01) | (0.01) | (0.01) | (0.01) | (0.01) | (0.01) |
| $\psi_{(d),1}$ | Origin y | 0.05 | 0.05 | 0.05 | 0.05 | 0.05 | 0.05 |
|  |  | (0.01) | (0.01) | (0.01) | (0.01) | (0.01) | (0.01) |
| $\psi_{(d),1}$ | Origin z | $-0.02$ | $-0.02$ | $-0.02$ | $-0.02$ | $-0.02$ | $-0.02$ |
|  |  | (0.00) | (0.00) | (0.00) | (0.00) | (0.00) | (0.00) |
| $\psi_{(a),1}$ | Destination x | $-0.00$ | $-0.00$ | $-0.00$ | $-0.00$ | $-0.00$ | $-0.00$ |
|  |  | (0.01) | (0.01) | (0.01) | (0.01) | (0.01) | (0.01) |
| $\psi_{(a),2}$ | Destination y | $-0.05$ | $-0.05$ | $-0.05$ | $-0.05$ | $-0.05$ | $-0.05$ |
|  |  | (0.01) | (0.01) | (0.01) | (0.01) | (0.01) | (0.01) |
| $\psi_{(a),3}$ | Destination z | 0.02 | 0.02 | 0.02 | 0.02 | 0.02 | 0.02 |
|  |  | (0.00) | (0.00) | (0.00) | (0.00) | (0.00) | (0.00) |
| Time (sec) |  | 215.55 | 273.09 | 219.23 | 379.66 | 217.04 | 378.72 |
| Iterations |  | 15 | 20 | 12 | 12 | 12 | 12 |
| Time/Iteration (sec) |  | 14.37 | 13.65 | 18.27 | 31.64 | 18.09 | 31.56 |



\end{document}